# Physics-informed Shadowgraph Network: An End-to-end Density Field Reconstruction Method

**Xutun Wang (王绪暾), Yuchen Zhang (张宇晨)\*, Zidong Li (李子栋),**
**Haocheng Wen (闻浩诚), and Bing Wang (王兵)\***

School of Aerospace Engineering Tsinghua University, Beijing 100084,

**Abstract:** This study presents a novel approach for quantificationally reconstructing density fields from shadowgraph images using physics-informed neural networks. The proposed method utilizes the shadowgraph technique visualizing the flow field, enabling reliable quantitative measurement of flow density fields. Compare to traditional methods, which obtain the distribution of physical quality in spatial coordinates case by case. We establish a new end-to-end network that directly from shadowgraph images to physical fields. Besides, the model employs a self-supervised learning approach, without any labeled data. Experimental validations across hot air jets, thermal plumes, and alcohol burner flames prove the model's accuracy and universality. This approach offers a non-invasive, real-time surrogate model for flow diagnostics. It is believed that this technique could cover and become a reliable tool in various scientific and engineering disciplines.

## 1. Introduction

Accurate measurement of the flow field, as a fundamental tool in various research topics, such as fluid mechanics, thermodynamics, etc., is still a challenging task to date[1]. It is true that the rapid evolution of digital imagery and the fast growth of computational power play a vital role in the development of image-based diagnostic techniques[2]. Of the developed techniques, particle image velocimetry (PIV) and planar laser-induced fluorescence (PLIF) could be an appropriate demonstration that most of those measurements required complicated and expansive setups[3]. The specified laser sources and artificial particles also prevent those "high-tech" methods from moving outside the laboratory condition. Thus, a simple, cost-effective, and real non-invasive technique to measure the flow field under common circumstances is worthwhile to develop.

By mentioning flow field measurement, there is no doubt that the velocity fields are one of the most basic and fundamental properties that deserve the dominant attention and efforts. For a general isothermal and incompressible (or weak compressible) condition, velocity is the majority of information that needs to be collected from the flow field. However, after taking the thermo or compress effect into account, one additional variable from the equation of state deserves to be studied in order to capture the thermodynamic effect. Usually, the obtainable information is density, pressure, and temperature[4]. The method introduced in this publication provides a brand-new means of measuring the density field.

Qualitative visualization of the density field has long been studied and used in the investigation of flow fields for the past few decades. There is a huge amount of methods that are able to achieve to goal of qualitative visualization, such as passive tracers of neutrally buoyant particles, shadowgraph, schlieren etc[5]. These methods are still popular in studying the phenomenon accompanied by aerodynamic or hydrodynamic effects. Researchers also keep attempts to extract quantitative information from qualitative visualization results for the past decades[6]. The early stage of investigation about quantitatively measuring the density fields focused mostly on the Schlieren technique. The idea of using the Schlieren method is straightforward since the method itself is mainly focused on the large density gradient[7]. And the sharp knife edge will enhance the change of greater density change. On the other hand, it will also neglect the less apparent part of the density field, which could explain the limitation of quantification of the density field based on the traditional Schlieren setup. Then, a novel method purely based on the computer vision technique of digital image correlation (DIC)[8] was proposed. Due to the highly developed image process program, the experiment setup has been simplified to an extreme extent, which only requires a background plane, a light source, and a camera Based on its fundamental theory of image processing, the method is called background-oriented schlieren (BOS). A huge number of researches have been undertaken following the rules of BOS[9–11]. However, due to its image processing fundament, this method is highly sensitive to spatial resolution due to its size of correlation box[12]. Since only a significant density variant could provide apparent image displacement, the BOS method is similar to the traditional Schlieren method in the aspect of focusing on a great density gradient.

The density diagnostic methods being introduced previously have been summarized in Table 1. By comparing various methods' properties, it is clear that the traditional Schlieren and Shadowgraph methods are not designed to quantitatively measure the density. While BOS designed to be able to quantitatively measurements. The result is sensitive to spatial resolution since it is purely based on computer vision algorithms to calculate the distortion[13,14]. To overcome the limitations described previously, the method introduced in this publication is based on the shadowgraph technique. The main difference between the shadowgraph and to Schlieren technique is the knife edge that cuts the strength of the light source. Without the knife edge, the gradient distribution will neither be enhanced nor diminished, which is similar to BOS in that only apparent displacement could be addressed from the source image. This could be a reliable starting point for quantitatively measuring the density of the flow field, as the shadowgraph method treated density variation equally instead of enhancing or diminishing part of the field. Previously, researchers used a numerical approach to solve the differential equations derived from the relationship between shadowgraph and density to solve simple cases such as pulsed positive streamer[15]. However, as the pre-solved equation is an elliptic partial differential equation, the result is highly sensitive to the boundary condition. Since determining the boundary condition in various scenarios is challenging this method is unlikely to be used for more general conditions. This research aims to establish an end-to-end direct mapping to infer the density field based on the Shadowgraph technique, without numerical solving partial equations. With the help of a recently developed deep learning method, the limitation of boundary condition sensitivity could be overcome.

Table 1. Comparison between various density diagnostic methods

| Method | Setups | Main Goal | Processing |
|---|---|---|---|
| **Schlieren** | Difficult | Qualitative | Real-time |
| **Shadowgraph** | Moderate | Qualitative | Real-time |
| **BOS** | Easy | Quantitative | Numerical Solving |
| **Ours** | Moderate | Quantitative | Directly Inference |

It has to be admitted that deep learning has achieved great progress in fields like computer vision and natural language processing. As a tool, the deep learning technique itself has been well developed, thus researchers have started to bend their attention to leveraging deep learning to model real-world physical problems and aid scientific discovery[16–18], including fluid mechanics[19]. It is true that traditional deep learning techniques, such as image processing, could address some challenges of physical problems like flow field reconstruction. But for more general and complicated scientific and engineering problems, the physical laws behind the phenomenon are not able to be neglected especially when compared with the solutions obtained from traditional methods. A key development in this direction is the introduction of physics-informed neural networks (PINNs) by Raissi et al.[20], which have become a powerful tool for solving complex problems. PINN has been developed rapidly in the past few years. Fluid dynamics is one of its focused topics[21–23]. As an important topic in the fluid dynamic study, the diagnostic of the flow field plays a vital role in the experimental investigation of fluid. Cai et al.'s work[9] could be a brief example to demonstrate the principle of the combination of PINN and traditional diagnostics techniques.

This study is grounded in practical optical physics. An end-to-end physics-informed shadowgraph network is proposed that utilizes shadowgraph images to reconstruct density fields. The model takes the shadowgraph images as inputs to reconstruct the refractive index field. Subsequently, the density field is derived from the refractive index field by the Gladstone-Dale law. Additionally, the model employs a self-supervised learning approach, making labeled data unnecessary. Experimental results demonstrate that the density field reconstruction method maintains high accuracy under various scenarios, including natural convection, forced convection, and combustion cases. As an efficient and precise reconstruction method, it serves as a promising surrogate model for real-time sensing and monitoring of thermal fluid. This approach not only enhances the understanding of fluid dynamics but also offers powerful non-invasive flow diagnostics tools, that are able to monitor and analyze in real-time.

This paper is developed as follows. Chapters 2 and 3 introduce the principle and implement details of the proposed method. Chapter 4 includes the reconstruction results and a discussion of serval cases. Finally, chapter 5 concludes the whole research.

## 2. Methodology

### 2.1 Shadow Imaging

The shadowgraph method is widely used in experimental fluid dynamics and heat transfer research. The main idea of this method is based on the principles of light refraction within inhomogeneous density fields. Figure 1 illustrates a typical schematic of the shadowgraph method. When light rays pass through a collimator, they form parallel rays that enter the experimental section. If there are disturbances in the experimental section caused by density variations, the light will be refracted to different extents based on changes in the refractive index. This leads to the formation of varying light-intensity images on the receiving screen. By observing and analyzing these images, information such as the refractive index in the experimental section can be obtained. The following is a brief derivation of the shadowgraph method.

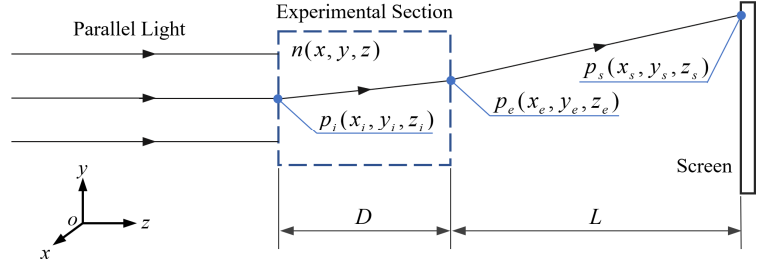

Fig. 1. Schematic of the shadowgraph method.

Firstly, it is assumed that the refractive index of the medium is determined by spatial coordinates $n = n(x, y, z)$. The coordinates of the incident point and the exit point of the light ray are represented as $p_i$ and $p_e$, respectively. The point where the light ray projects onto the screen can be expressed as follows.

$$x_s = x_e + L \cdot \frac{dx}{dz}\Big|_{p_e}$$
$$y_s = y_e + L \cdot \frac{dy}{dz}\Big|_{p_e}$$
(2.1)

Equation (2.1) can be further simplified by the weak refraction assumption[24].

$$x_s = x_i + L \cdot \int_{z_i}^{z_e} \frac{\partial \log n}{\partial x} dz$$
$$y_s = y_i + L \cdot \int_{z_i}^{z_e} \frac{\partial \log n}{\partial y} dz \qquad (2.2)$$

This assumption posits that the light rays that strike the entrance plane perpendicularly experience only minor deflections within the uneven field, yet they exhibit a measurable curvature upon exiting the experimental setup.

To further analyze the light rays projected on the screen. After passing through the experimental region with a non-uniform refractive field, the light intensity of the light source $I_0(x_i, y_i)$ and screen $I_s(x_s, y_s)$ can be expressed as follows.

$$I_s(x_s, y_s) = \sum_{(x_i, y_i)} \frac{I_0(x_i, y_i)}{\left|\frac{\partial(x_s, y_s)}{\partial(x_i, y_i)}\right|} \qquad (2.3)$$

Where $\left|\frac{\partial(x_s, y_s)}{\partial(x_i, y_i)}\right|$ is the Jacobian determinant for the transformation of the coordinate system from $(x_i, y_i)$ to $(x_s, y_s)$.

Due to the infinitesimal displacement assumption, the higher-order terms in the Jacobian determinant can be neglected. Thus, the mapping function can be approximated as a linear transformation, and the Jacobian determinant can be expressed as follows.

$$\left|\frac{\partial(x_s, y_s)}{\partial(x_i, y_i)}\right| = 1 + \frac{\partial(x_s - x_i)}{\partial x} + \frac{\partial(y_s - y_i)}{\partial y} \qquad (2.4)$$

So, the Eq. (2.3) can be rewritten as follows.

$$I_s(x_s, y_s)\left[1 + \frac{\partial(x_s - x_i)}{\partial x} + \frac{\partial(y_s - y_i)}{\partial y}\right] = \sum_{(x_i, y_i)} I_0(x_i, y_i) \qquad (2.5)$$

Finally, substituted Eq. (2.2) into Eq. (2.5) and integrated over the dimensions of the experimental section. The relationship between light intensity variation in the shadowgraph image to the refractive index field can be modeled using a Poisson equation as follows.

$$\frac{I_0(x_i, y_i) - I_s(x_s, y_s)}{I_s(x_s, y_s)} = (L \times D)\left(\frac{\partial^2 \log n(x, y)}{\partial x^2} + \frac{\partial^2 \log n(x, y)}{\partial y^2}\right) \qquad (2.6)$$

For transparent media, refractive index techniques rely on the specific relationship between refractive index and density, known as the Lorentz–Lorenz formula, which is expressed as follows[25].

$$\frac{n^2-1}{\rho(n^2+2)} = \text{constant} \tag{2.7}$$

Where $n$ is the refractive index and $\rho$ is the density. For gases such as air, where the refractive index is close to unity, this relationship simplifies to the Gladstone–Dale equation[25].

$$\frac{n-1}{\rho} = \text{constant} \tag{2.8}$$

The above equation establishes the relationship between the refractive index field and the density field. In measurements, the left side of Eq. (2.6) is recorded as the shadowgraph image. By solving this equation and using the Gladstone–Dale relation, the refractive index field and density field can be obtained.

## 2.2 Physics-Informed Density Field Reconstruction

For solving Eq. (2.6), numerical methods have been applied (such as finite difference method (FDM), and finite volume method (FVM) to solve the density field on spatial coordinates $n(x,y)$. There are two main shortages of this kind of method. First, due to the complexity of real problems, ideal boundary conditions such as Dirichlet and Neumann conditions do not exist, which may introduce errors in the result. Second, numerical methods can only obtain the solution $n(x,y)$ for a single case at a time and are relatively time-consuming. For each case, the result could only be obtained after convergence, which is unlikely to be highly efficient. Honestly, recently developed deep-learning-based methods handle the first shortages. Such as the study of He et al.[26], which uses the methods of neural radiance fields[27] in computer vision to construct an implicit function of the spatial distribution of $n(x,y,z)$. However, the second shortage still exists even with the help of this method.

In this research, we proposed a novel method that effectively addresses the two aforementioned limitations. Leveraging recent advancements in deep learning, many complex mapping relationships can now be modeled using the powerful tool of neural networks. The proposed method constructs an end-to-end mapping function that directly transforms the shadowgraph image light intensity field into the corresponding density field.

$$\rho = F_{map}(I) \tag{2.9}$$

Where $\rho$ is the density field, $I$ is the light intensity field from shadowgraph, $F_{map}(\cdot)$ is the end-to-end reconstruction mapping function. According to the universal approximation theorem[28], a neural network $NN(\cdot,\theta)$ with parameters $\theta$ can be used to approximate the mapping function $F_{map}(\cdot)$. In this way, we construct a new mapping relationship instead of directly solving the Poisson equation and the spatial distribution of the refractive index. The specific implementation details are provided as follows.

The simplified equation mentioned in the previous section can be further processed, as Eq. (2.1),

$$I_{s,norm} = \frac{\partial^2 \log n}{\partial x^2} + \frac{\partial^2 \log n}{\partial y^2} \qquad (2.10)$$

Where the left-hand side term $I_{s,norm} = (I_s - I_0)/(I_0 \times L \times D)$ represents the normalized light intensity data. The right-hand side of the equation consists entirely of the second-order derivative of the logarithmic refractive index $\log n$. Given the known light intensity $I_{s,norm}$, the refractive index field can be obtained by solving this Poisson equation. From the refractive index field, the density field can then be derived using the Gladstone-Dale relation. However, solving this Poisson equation is not a typical forward problem, as there is no explicit expression for the boundary conditions. This is where PINNs offer an advantage in solving inverse problems[20].

PINNs followed the universal approximation theorem[28] of neural networks to approximate the solution function of a differential equation, mapping inputs to corresponding outputs. During this training process, the derivative terms are constructed using automatic differentiation[29] and incorporated into the loss function. Boundary or initial conditions are also included. Once the training converges, the network's predictions satisfy the physical constraints, thereby providing an approximate solution to the original problem. Due to the complexity of image processing, this method did not use the fully connected neural network with automatic differentiation. So, the proposed method is a type of 'physics-informed' neural network, but it's different from the typical 'PINN'. Overall, this approach fits into the broader field of physics-informed deep learning[30].

This physics-informed shadowgraph network is based on a convolutional neural network (CNN)[31], where both the input and output are single-channel tensors. A schematic of the model is shown in Fig. 2. The model adopts an encoder-decoder architecture, comprising one encoder and two separate decoders. As shown in Fig. 2(a), the encoder is responsible for extracting latent features from the normalized light intensity data. The first decoder, referred to as the Image Reconstruction (IR) decoder, reconstructs the original image from these features. This reconstruction process facilitates the encoder in learning more meaningful representations from the input data. As depicted in Fig. 2(b), the second decoder, named the Refractive Index (RI) decoder, reconstructs the refractive index field using the features extracted by the trained encoder. This reconstruction is further constrained by incorporating physical laws into the training process, ensuring that the resulting refractive index field adheres to the underlying physical principles.

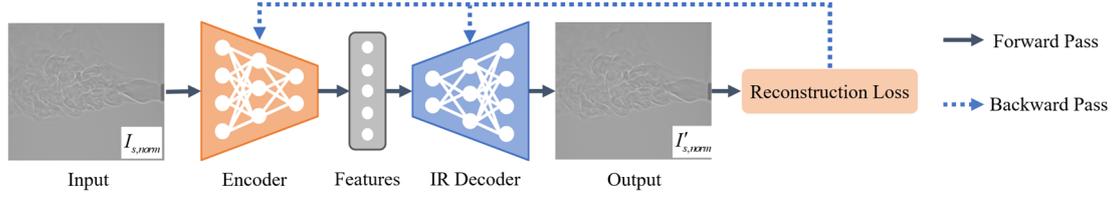

(a). The encoder with IR decoder for image reconstruction

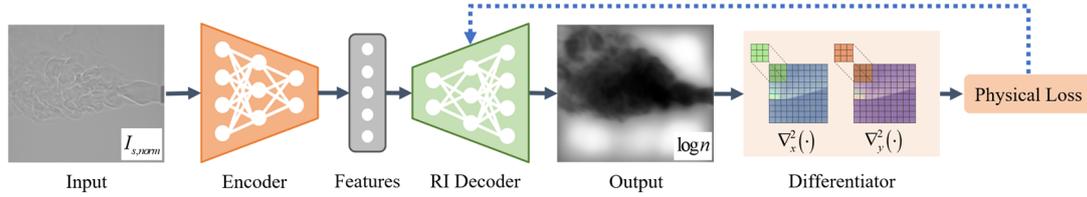

(b). The encoder with RI decoder for refractive index field reconstruction

Fig. 2. Schematic of the proposed neural network structure.

For the input tensor $I_{s,norm}$ of the encoder, by subtracting $I_0$, the influence of dust on the camera image sensor and vignetting effects are mitigated. Subsequently, normalizing the intensity by dividing $I_0$ eliminates the influence of the light source's intensity. Considering that the captured images exhibit brighter and darker regions compared to the background, the results contain both positive and negative values. Therefore, a transformation is necessary to map these values within a suitable range for processing by the neural network. This study directly employs the sigmoid function as the input transformation. This transformation also maintains monotonicity. After the transformation, the tensor fed into the encoder is normalized within the range of 0 to 1. In fact, this operation also enhances the model's generalization performance[32].

For the output layer, since the refractive index of air is approximately 1.00027, the changes in refractive index due to density variations occur at the order of $10^4$. This small magnitude is disadvantageous for the neural network's output. Therefore, the output of the RI decoder is transformed to $10^3 \log n$, ensuring that the output values are within a more reasonable range.

The implementation of these control equations in neural networks depends on the construction of partial differential terms. As shown in Fig. 3 and Eq. (2.3), the physical loss is calculated with a normalized light intensity field and spatial derivatives of the logarithmic refractive index field. The spatial derivatives are calculated by the finite difference convolution filter[33,34].

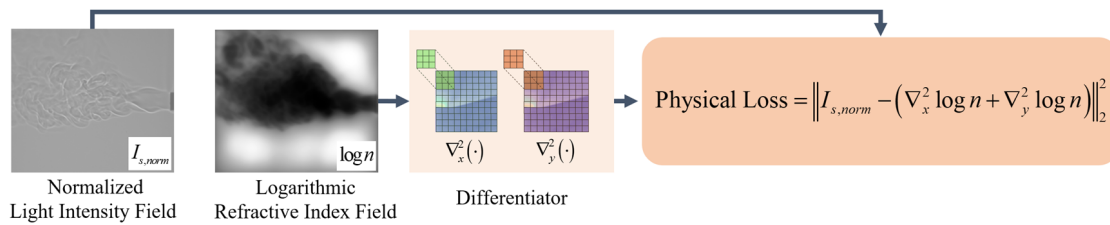

Fig. 3. Schematic of physical loss calculation.

$$L_p(\theta_{RI\ decoder}) = \left\| I_{s,norm} - \left(\nabla_x^2 \log n + \nabla_y^2 \log n\right) \right\|_2^2 \qquad (2.11)$$

Where $L_p$ describes the physical loss of the RI decoder, $\theta_{RI\ decoder}$ is the trainable parameter of the RI decoder. The above loss function is the MSE (Mean Squared Error) loss. However, in this research, the smooth L1 loss performs better than the MSE loss and L1 loss.

## 3. Problem Setup

In this chapter, an introduction in detail about this study will be given. First, the implementation details of the shadowgraph technique that is being utilized in this research will be recapped. Then, the training details about the neural networks will be discussed.

### 3.1 Experiment Setup

The schematic experimental setup of the shadowgraph method is shown in Fig. 4. In this research, dual concave mirrors are adapted to produce high-quality visualizations of refractive index variations[24]. A point light source, formed by emitting light passing through a narrow slit, is placed at the first concave mirror's focus point. The light is then reflected by the first concave mirror, which transforms the diverging rays into a parallel light field. These parallel rays of light travel through the experimental section. The experimental section is located at the center of the optical path, where the target device or phenomenon is placed for observation. The second concave mirror gathers the transmitted parallel light and focuses it towards the camera sensor, ensuring that any perturbations in the light path, caused by changes in refractive index in the experimental section, are captured with high fidelity.

The camera, positioned behind the focal point of the second concave mirror, records the intensity variations produced by these refractive index changes. The measurement section observed by the camera is approximately 110 × 88 mm, allowing for a precise field of view within which the experiment is conducted.

The setup of dual concave mirrors ensures that the distortions are minimized. By using concave mirrors for both collimation and focusing, the design minimizes optical aberrations and ensures that subtle variations in the refractive index within the experimental section can be accurately captured.

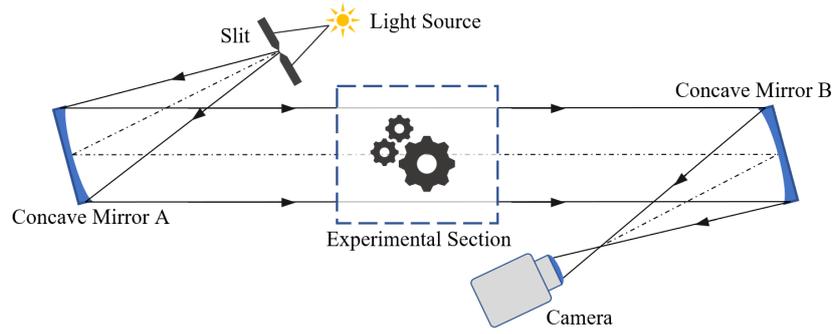

Fig. 4. The schematic of the shadowgraph experiment setup

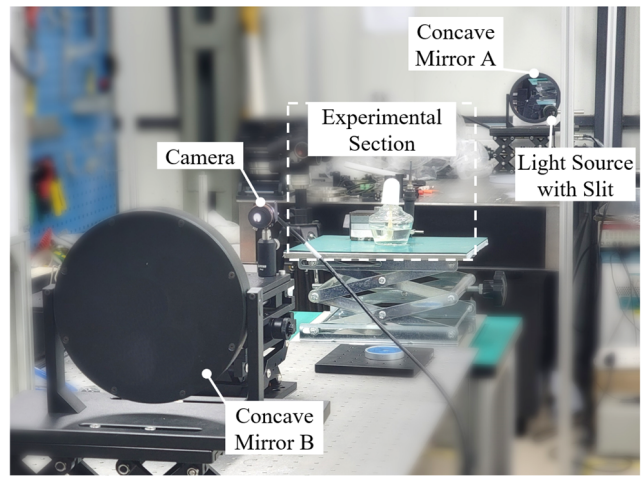

Fig. 5. Photo of shadowgraph experiment equipment

## 3.2 Self-supervised Training Setup

The training process of the model consists of three steps: pre-training the encoder, training the RI decoder, and fine-tuning the RI decoder. The model employs a self-supervised learning approach, without any labeled data. Since this research is not supervised learning, all training and testing exist on a single dataset. The proposed neural network model is implemented by the open-source deep-learning framework PyTorch in Python language. The source code can be found in the following link[1].

1. **Pre-training the Encoder**

The encoder is the foundation part of the model, which is used to extract features from the original light intensity field. The quality of the features directly influences the performance of the subsequent decoders. High-quality features enhance the model's ability to reconstruct the refractive index field. However, the encoder cannot be directly trained, as its output features lack corresponding labels. Therefore, a self-supervised training approach is adopted. This involves constructing a decoder following the encoder to reconstruct the input itself.

---

[1] https://github.com/pyrimidine/Physics-informed-Shadowgraph-Density-Field-Reconstruction.git

$$I'_{s,norm} = NN\left(NN\left(I_{s,norm}, \theta_{encoder}\right), \theta_{IR\ decoder}\right) \tag{3.1}$$

Where $\theta_{encoder}, \theta_{IR\ decoder}$ are the network parameters of the encoder and the IR decoder, respectively. $I'_{s,norm}$ is the output of the IR decoder, which is aimed to be equal to $I_{s,norm}$.

The loss function of this pre-training process $L_{pre-training}$ can be written as follows. The $L_{reconstruction}$ describes the self-supervised reconstruction loss of the encoder and IR decoder. The reconstruction loss is used to evaluate the accuracy of the outputs compared to the inputs.

$$L_{pre-training} = L_{reconstruction}(\theta_{encoder}, \theta_{IR\ decoder}) = \left\|I_{s,norm} - I'_{s,norm}\right\|_2^2 \tag{3.2}$$

**2. Self-supervised Training of RI Decoder**

The RI decoder is the crucial part of the model. The reconstruction process of the refractive index field can be written as follows.

$$n = NN\left(NN\left(I_{s,norm}, \theta_{encoder}\right), \theta_{RI\ decoder}\right) \tag{3.3}$$

Where $\theta_{RI\ decoder}$ is the network parameters of the RI decoder. $n$ is the refractive index field, which is regularized by the governing equation. The RI decoder is trained totally unsupervised in this study.

The loss function of the RI decoder training process $L_{training}$ consists of two parts, physical loss $L_{physical}$ and reference loss $L_{reference}$. The physical loss has been discussed before. The reference loss is used to calibrate the output of the RI decoder by comparing it with a reference field with a known refractive index.

$$L_{reference} = \left\|n - n_0\right\|_2^2 \tag{3.4}$$

The whole loss of the RI training is the weighted sum of physical loss and reference loss. In this study the $w_p$ is 1 and $w_r$ is set as 10 to balance the magnitude of different loss items.

$$L_{training} = w_r L_{reference} + w_p L_{physical} \tag{3.5}$$

**3. Fine-tuning the RI Decoder**

Although the model trained above can be analyzed qualitatively, further fine-tuning of the RI decoder is necessary to achieve quantitatively accurate results. In addition to the reference loss and physical loss, the fine-tuning process introduces a value loss, which constrains the magnitude of the output. The value loss can take various forms, such as limiting the minimum and maximum values of the measured flow field. The final loss function is the weighted sum of these three components.

$$L_{fine-tuning} = w_v L_{value} + w_r L_{reference} + w_p L_{physical} \tag{3.6}$$

# 4. Result and Discussion

In this chapter, further analysis based on the data collected from the experiments will be undertaken.

## 4.1 Horizontal Hot Air Jet

Hot air jets are used for many purposes and have been widely used in almost every aspect, from house heaters to polishing, from hair dryers to jet engines. Therefore, this condition could be a great case to demonstrate the utilization of this method. In this study, the horizontal hot air jet is chosen to be the study case.

In the experimental setup, a hot air gun was positioned to the left of the measurement region, with its nozzle directed toward the area of interest. The temperature of the hot air gun can be adjusted between 100°C and 500°C, and the outlet air velocity is 1.6 m/s. Fig. 6 shows an example of the raw image and normalized image. The normalized image effectively removes noise caused by dust and the influence of the nozzle, resulting in improved contrast. The normalized image is distributed more evenly between 0 and 1. This makes better use of the available grayscale values for further processing.

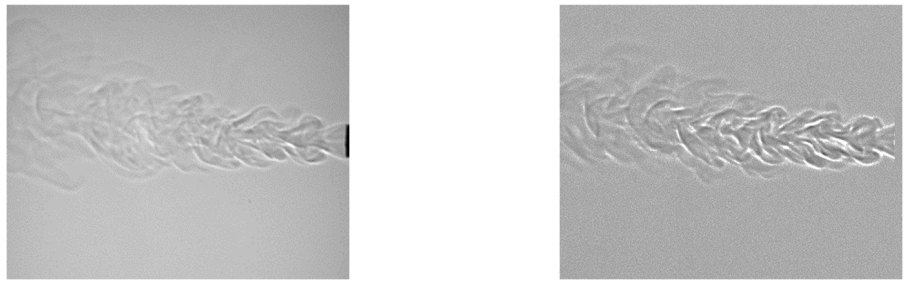

Fig. 6. One example of the raw image (left) and normalized image (right).

The following Fig. 7 demonstrates the reconstructed density and temperature fields. The density calculation applies the GD relation, where the GD constant of air is $2.24 \times 10^{-4}$. In this experiment, the data of the density field has been collected to study the phenomenon. In order to validate the reconstruction results with the thermocouple test, the temperature field was calculated based on the ideal gas law, assuming constant pressure at 1 atm and using the gas constant for air $287 \, \text{J} \cdot \text{K}^{-1} \cdot \text{kg}^{-1}$. Under these conditions, the temperature was derived from the reconstructed density field, allowing for direct comparison with experimental measurements.

The reconstructed results well capture the details of the flow field, including the jet core, the uplift of the jet influenced by buoyancy, and the diffusion process. A higher temperature usually comes with higher gas viscosity, resulting in a larger jet expansion angle and eventually leading to a more complicated flow field. Besides, due to the property of the air of a higher temperature also comes with a lower density, temperate effects on buoyancy need to be taken into consideration. It is important to note that this study overlooks three-dimensional effects. As temperature increases, the jet expansion angle also widens, occupying a greater thickness of jet flow. In this study, the thickness is assumed to be fixed, which results in the high-temperature region not being located at the nozzle exit. Future research will address corrections related to three-dimensional effects.

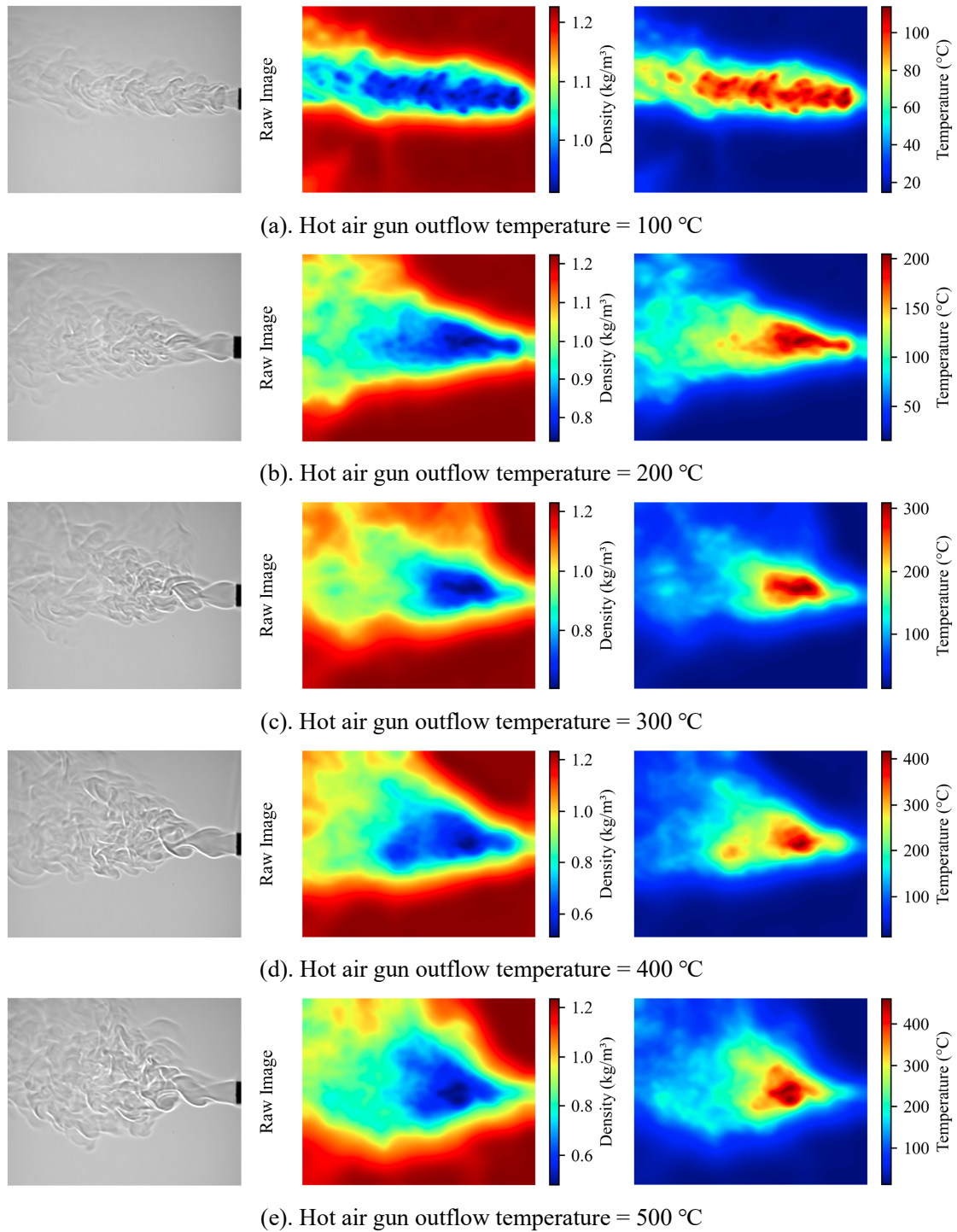

(a). Hot air gun outflow temperature = 100 °C

(b). Hot air gun outflow temperature = 200 °C

(c). Hot air gun outflow temperature = 300 °C

(d). Hot air gun outflow temperature = 400 °C

(e). Hot air gun outflow temperature = 500 °C

Fig. 7. Raw image (left), reconstruction density field (middle), and temperature field (right) at different hot air gun outflow temperatures.

In this study, a thermocouple was deployed 10 mm from the nozzle exit to acquire the time-averaged temperature of the jet core. The reconstructed temperatures of the jet core are compared to experiment measurements are shown in Fig. 8. Each box represents the reconstructed values at 300 different time instances. The average values of the reconstructed results show good agreement with the experimental measurements, with a relative error of less than 5%.

Besides, it can be found that the variance of the jet core temperature increases with the rise of the setting temperature. This could be explained by that a stronger turbulence effect usually comes with a higher temperature. The stronger turbulence effect brings a strong stochastic effect which results in the greater variance in the distribution. It will also lead to an inaccurate reconstruction result. Nevertheless, the model still provides reasonable results under this condition.

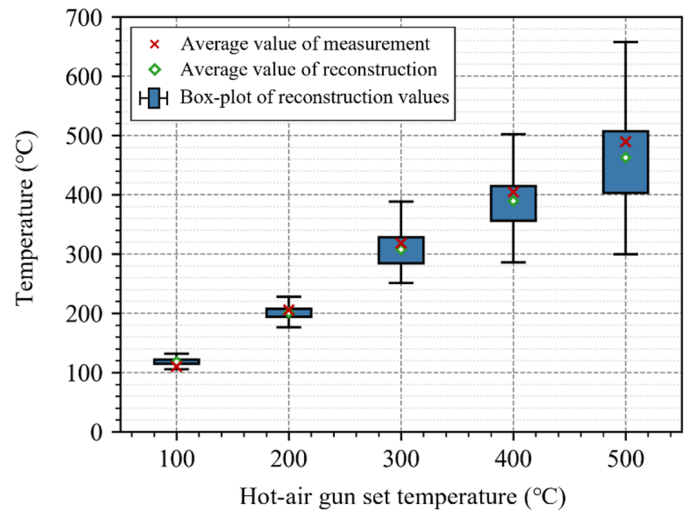

Fig. 8. Comparison of reconstructed temperature values at 300 different time instances with experimental measurements.

Overall, the method proposed in this study effectively reconstructs the density fields of the hot air jet flow. It captures the instantaneous flow structure of the jet flow field and demonstrates good agreement with the experimental measurements.

## 4.2 Thermal Plume

Thermal plumes, as the most basic form of thermal effect on fluid dynamics, could be easily observed across over the world. All types of natural convection flow could be assumed as a special type of thermal plume. Therefore, it is worthwhile to use a thermal plume to study the universality of the proposed method. In this case, the electric soldering iron tip is utilized to study the thermal plume's flow field.

In the experimental setup, an electric soldering iron tip is placed at the lower end of the measurement area to ensure that the natural convection thermal plume is sufficiently strong and can be accurately measured. The temperature of the soldering iron tip is set to 500 °C. Figure 8 shows the captured raw image along with the reconstructed flow field results. It is evident that the model reconstructs reasonable density distributions of the plume.

It is important to note that the background is not ideally uniform due to heat dissipation from other components of the heating device; however, this does not affect the results for the plume core. Additionally, the soldering iron emits a faint red visible light at this temperature. However, since the intensity of the background light source is sufficiently high, its grayscale value in the raw image is approximately two orders lower than the background, making its influence negligible.

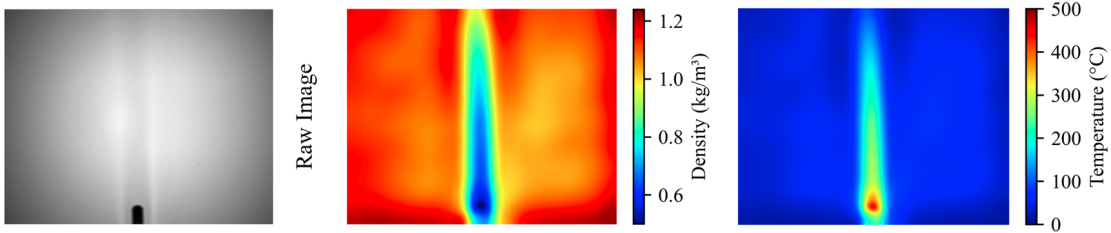

Fig. 9. Raw image (left), reconstruction density field (middle), and temperature field (right) of soldering iron thermal plume.

Figure 9 shows the average flow field from 500 cases. All coordinates were normalized using the tip diameter $d_0$. The plume is symmetrically distributed along the axis, and the flow structure is clearly visible. Including plume flow structures, such as the entrainment of surrounding air, and the diffusion of the plume boundary. As mentioned above, the background is not ideally uniform due to heat from other components outside the view. Besides, a slight leftward drift is observed at the top of the plume. This is due to the weak nature of natural convection makes it highly sensitive to external disturbances. A quantitative analysis of the reconstructed results will be discussed below. As mentioned above, the temperature field was also calculated from the ideal gas equation to quantitatively evaluate the model.

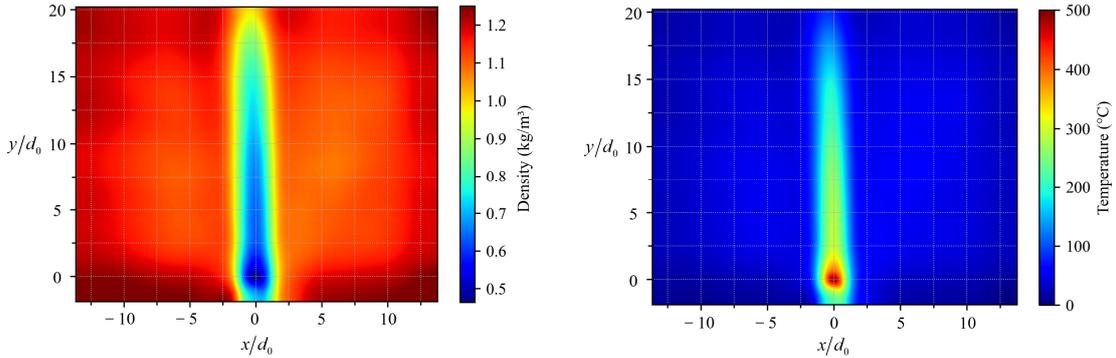

Fig. 10. Average density field (left) and temperature field (right) of the plume.

Figure 11 shows the temperature distribution at $x = 0$, which is the center line of the plume. Five thermocouples were arranged each 1.25 $d_0$ from the iron tip to acquire the time-averaged temperature from the plume core. The figure shows that the reconstructed results align with the experiment measurements at the beginning. Starting from $y = 2.5$ $d_0$, the values showed a little difference but the overall trend still remains reasonable. Meanwhile, the standard deviation also increases from that point. By comparing with the input raw image, it can be found that the deviation is primarily due to the small gradient of light intensity in this region. Since the camera only has a limited bit depth, it may not capture minor changes in light intensity. It will directly lead to insufficient precision in solving the governing equations and cause inaccurate results. Additionally, the plume is highly sensitive to external disturbances, often exhibiting oscillations during the imaging process, which makes it more challenging to reconstruct accurately. We also used the commercial CFD software FLUENT to simulate this case, obtaining the same conclusion. Since the thermocouples are contact sensors of measurement, their impact on the flow field cannot able to be neglected. This could explain why its results are lower compared to those of the CFD simulation and reconstruction results.

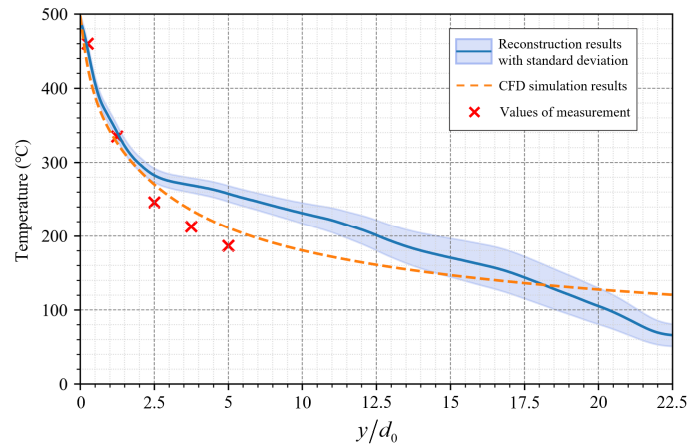

Fig. 11. Temperature distribution on reconstruction results with strand deviation, CFD simulation results, and experiment measurement values on the center line of the plume.

For quantitative comparison, the temperature distribution has been analyzed at $y = 0.25\ d_0$, $y = 1.25\ d_0$, and $y = 2.5\ d_0$ in Fig. 12. The figure shows that the reconstructed results align closely with the CFD simulation. However, as the plume rises, a slight leftward shift appears due to the interference from the environment.

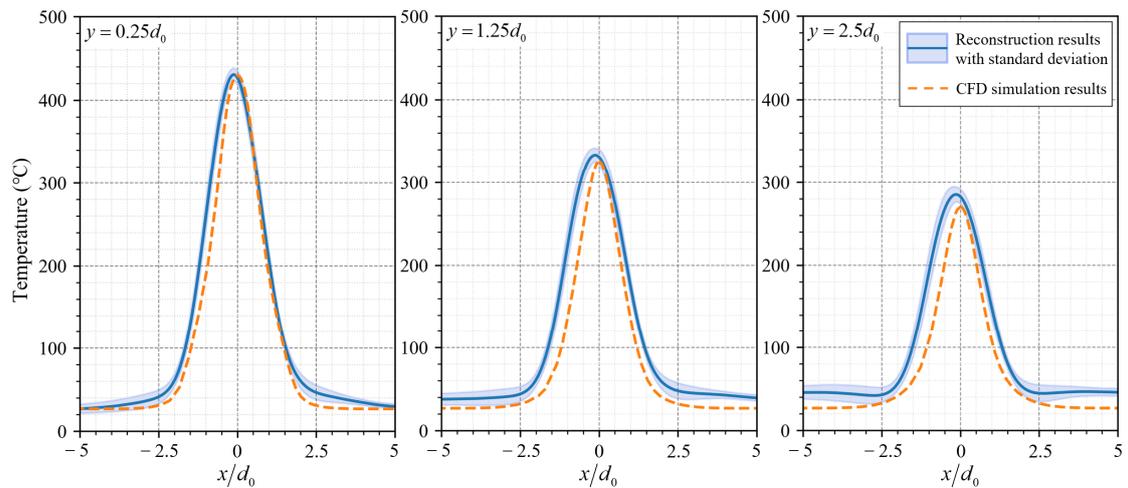

Fig. 12. Temperature distribution of reconstruction results with strand deviation and CFD simulation results on $y = 0.25\ d_0$ (left), $y = 1.25\ d_0$ (middle), and $y = 2.5\ d_0$ (right).

Overall, the method proposed in this study successfully reconstructs the flow fields of the plume flow. It accurately captures the flow structures of the plume and shows strong consistency with the experimental observations and CFD results.

## 4.3 Alcohol Burner Flame

The alcohol burner is a piece of laboratory equipment used to produce an open flame. Due to its reliability and widespread use, it could be a perfect study case to demonstrate the availability of the proposed method on a flow field with multiple species and chemical reactions, which are quite common in the thermal fluid field.

In the experimental setup, this study arranged a vertical lift platform, on which an alcohol burner was ignited. By moving the vertical lift, the imaging area was adjusted to focus on either the flame core or the flame plume. The high background brightness ensured that the luminosity of the alcohol flame did not affect the measurement results. However, due to the difficulty in quantitatively describing the combustion process of the alcohol burner (such as the shape of the wick), only qualitative and semi-quantitative reconstruction results related to the alcohol burner, such as flame structure, are presented. This research aims to demonstrate the feasibility of the method and to validate it through several standardized cases, with more detailed investigations into the flame structure planned for the future.

Fig. 13 and Fig. 14 present a series of reconstruction results at various time steps for the wick region of the alcohol burner flame and plume, respectively. The results clearly capture the inner flame, outer flame, and plume of the alcohol flame, as well as the oscillation process of unstable combustion. The reconstructed density and temperature fields semi-quantitatively reflect the flame structure, providing more and clearer information than the original shadow images. A video of the reconstruction results obtained by this method can be found in the following link[2]. With a single NVIDIA GeForce RTX 4090 GPU, the forward prediction speed is on the order of 10 ms for each frame. The link also showcases the real-time reconstruction results using a laptop. However, due to the limited computational power, the reconstruction results experienced some delays.

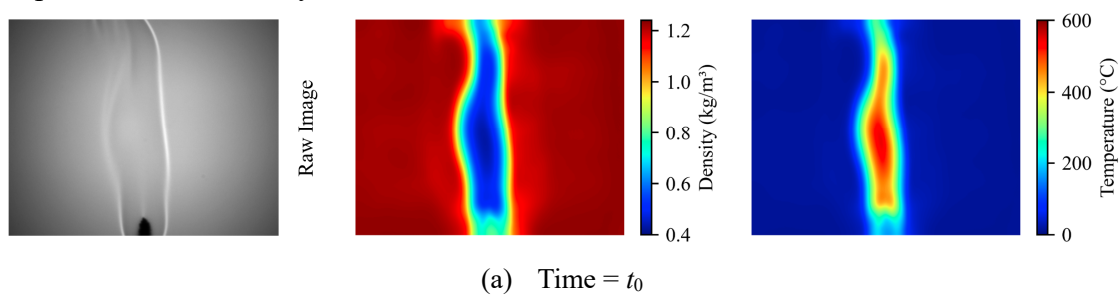

(a)  Time = $t_0$

---

[2] https://github.com/pyrimidine/Physics-informed-Shadowgraph-Density-Field-Reconstruction/tree/f593a1f407dd9a37929c4ccc6eb1536b537d3ba6/results

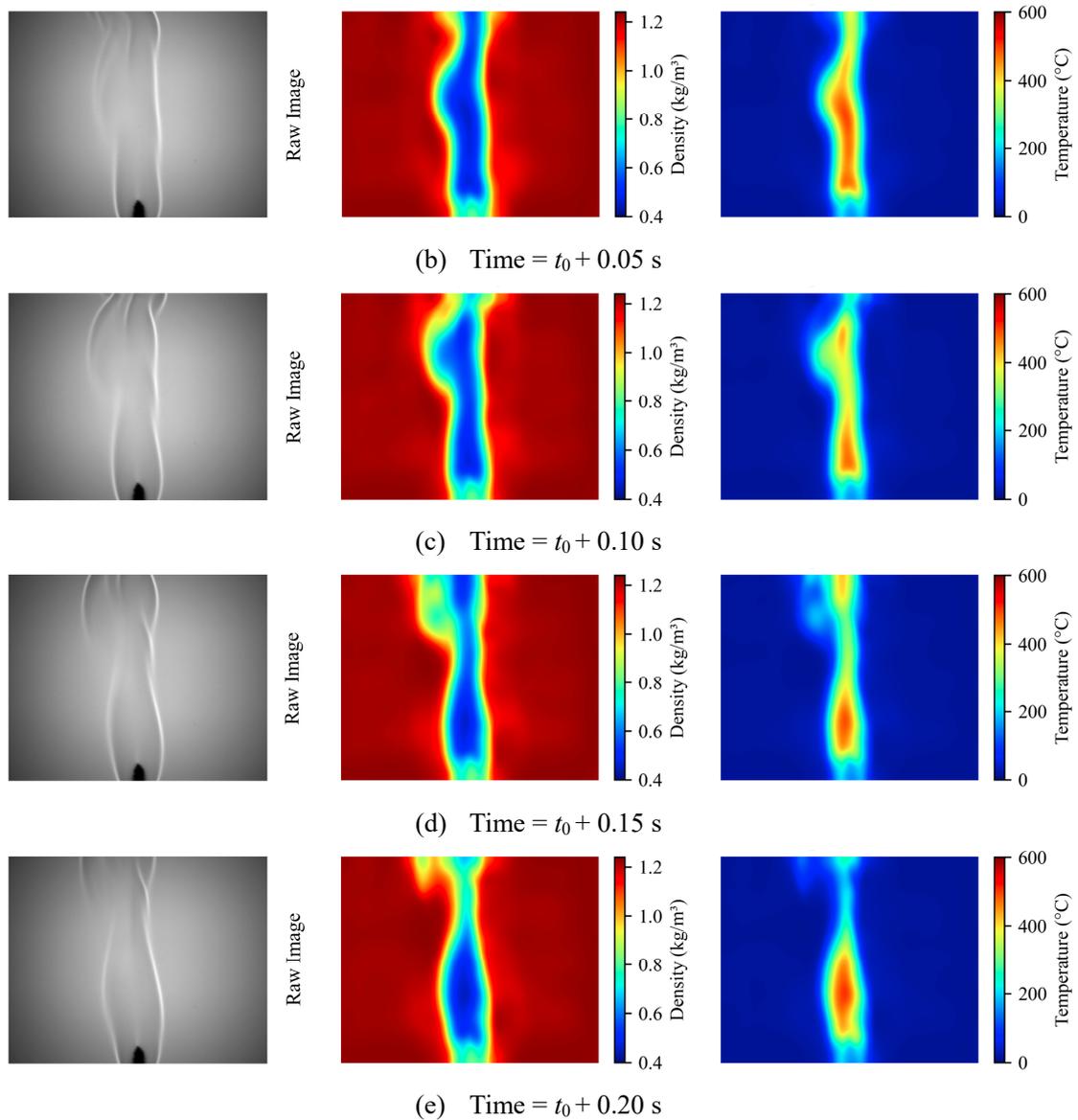

(b) Time = $t_0 + 0.05$ s

(c) Time = $t_0 + 0.10$ s

(d) Time = $t_0 + 0.15$ s

(e) Time = $t_0 + 0.20$ s

Fig. 13. Raw image (left), reconstruction density field (middle), and temperature field (right) of alcohol burner flame core at different a series of time instances.

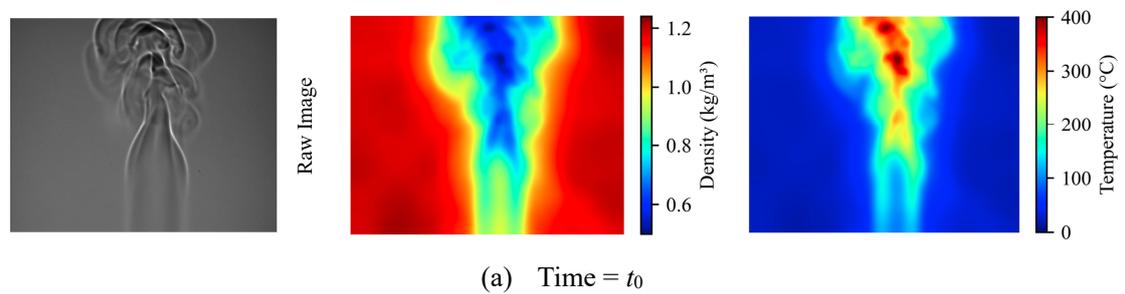

(a) Time = $t_0$

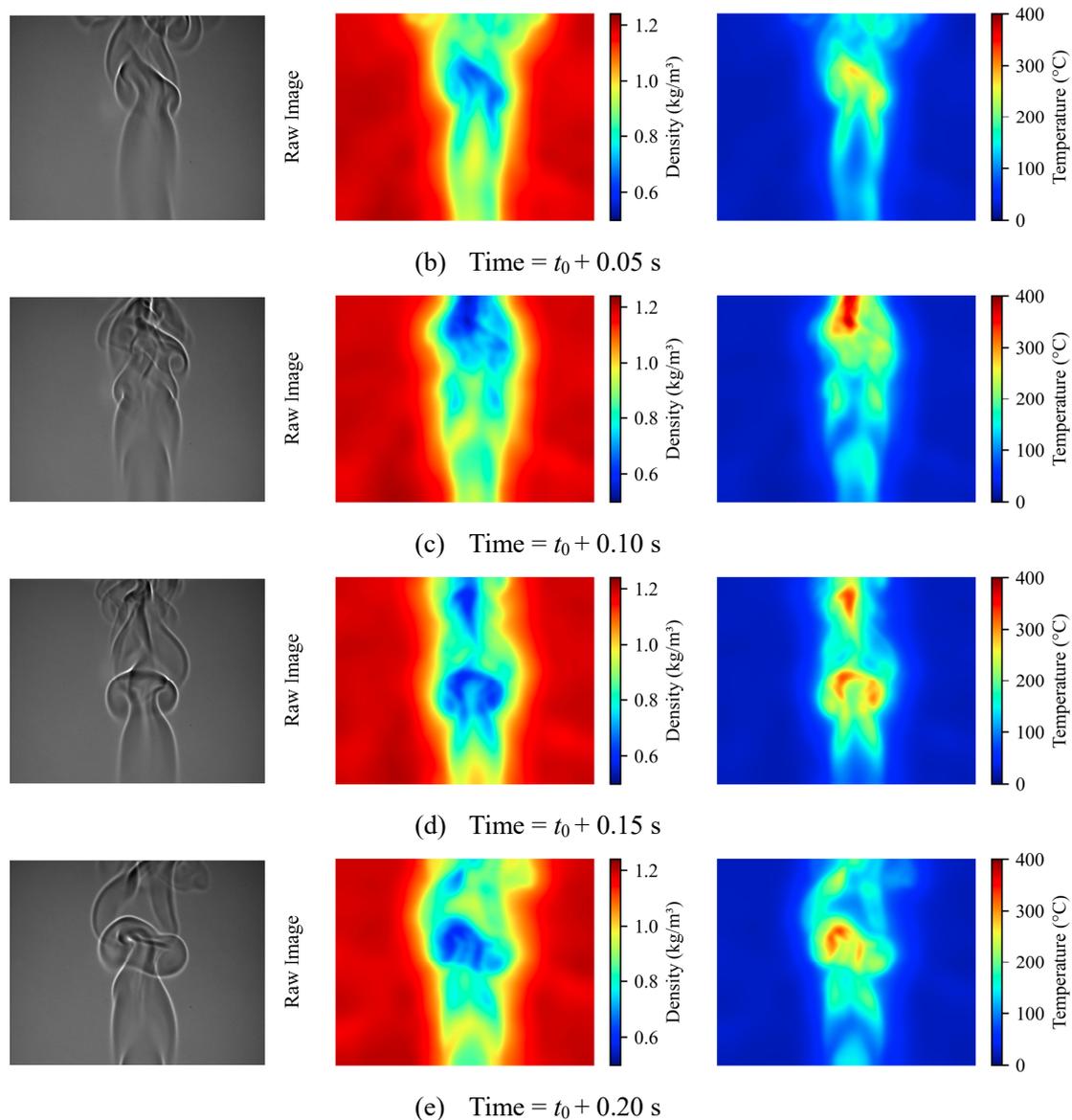

(b) Time = $t_0$ + 0.05 s

(c) Time = $t_0$ + 0.10 s

(d) Time = $t_0$ + 0.15 s

(e) Time = $t_0$ + 0.20 s

Fig. 14. Raw image (left), reconstruction density field (middle), and temperature field (right) of alcohol burner flame plume at different a series of time instances.

Overall, the test cases for alcohol burners have been quantitatively analyzed by the proposed method. After reviewing the reconstruction result, the flame structures have been discussed in detail, which provides a shred of evidence about the capability of the method in analyzing the flow field coupled with multispecies and chemical reaction phenomena.

## 4.4 Compared with BOS

In this section, we compared the results of our method with those of the BOS technique. The BOS method relies on solving for the displacement of the background, with its calculations dependent on a cross-correlation algorithm commonly used in Particle Image Velocimetry (PIV)[35]. The processing programming is based on the open-source framework OpenPIV[36]. Different correlation box sizes have been implemented into compression, which concluded that the proposed method's overall performance is compatible with the BOS method. Additionally, the computational cost and efficiency work better than that of the BOS method. Detailed results can be found in Fig. 15 and Fig. 16.

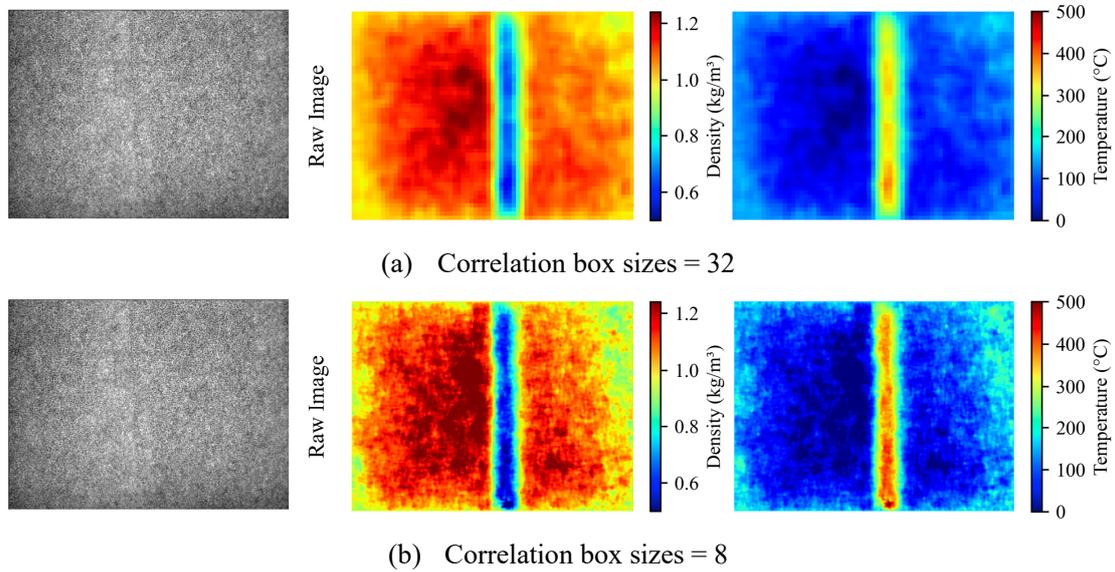

(a) Correlation box sizes = 32

(b) Correlation box sizes = 8

Fig. 15. Raw image (left), reconstruction density field (middle), and temperature field (right) of soldering iron thermal plume with different correlation box sizes.

Figure 15 shows the results of using the BOS method to analyze the thermal plume from the soldering iron, with the soldering iron tip positioned below the bottom boundary. The results closely resemble those mentioned earlier; however, there are some discrepancies at the edges and in areas with steep gradients. By adjusting the correlation box size, we found that a larger box reduces noise but results in lower resolution in the reconstructed outcomes.

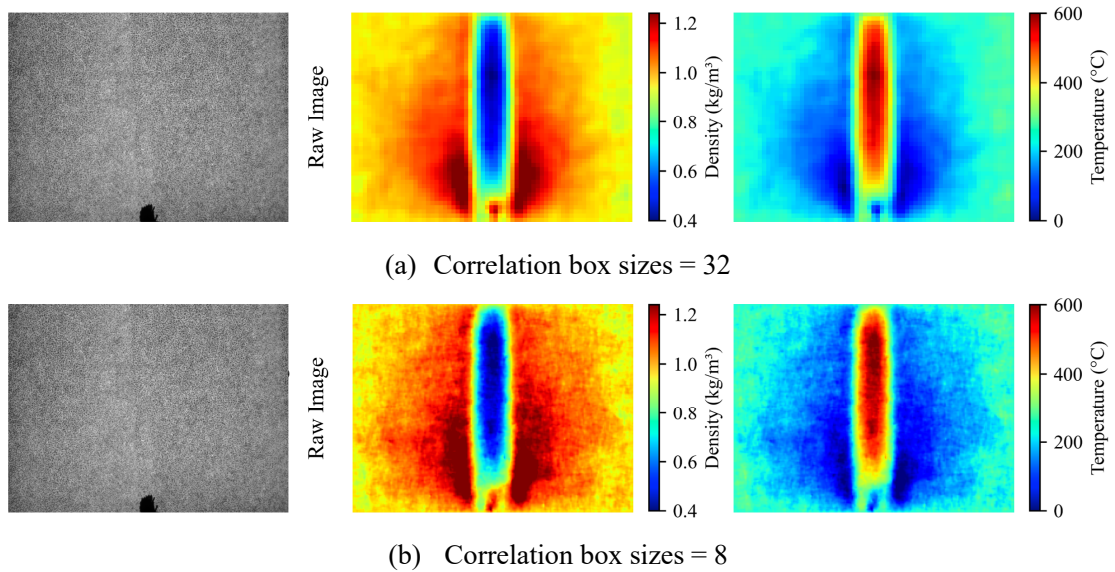

(a) Correlation box sizes = 32

(b) Correlation box sizes = 8

Fig. 16. Raw image (left), reconstruction density field (middle), and temperature field (right) of alcohol burner flame with different correlation box sizes.

Figure 16 shows the results of using BOS to analyze the alcohol burner flame. Compared to Section 4.3, using the BOS method with a smaller correlation box results in higher noise, while a larger box reduces noise but leads to lower resolution in the reconstructed results.

After comparing the density field that is reconstructed by using the proposed method and BOS, it could be concluded that both methods could provide a promising estimation of the density field. However, the proposed method performs better in capturing detailed features. By considering the processing manner, the proposed method is based on direct inference but the BOS is based on an iterative method. In summary, this method offers advantages in both the quality and speed of the density field reconstruction. The advantages of this method can be mainly summarized as high fidelity and low processing time.

## 5. Conclusion and future works

This study presents a novel method of density field reconstruction, called physics-informed shadowgraph network. This method integrates physical principles and employs deep learning techniques to construct an end-to-end self-supervised model that directly utilizes shadowgraph images to reconstruct density fields without any labeled data. The proposed technique offers a non-invasive, real-time sensing alternative to traditional flow field measurement methods.

Three different test cases have been utilized to study the versatility and universality of the proposed method. The results from the three different scenarios obtain reasonable results. In order to validate them, thermocouple tests have been undertaken in both hot air jet and thermal plume cases. Due to the limitation of the contact test by using the thermocouple, A CFD simulation has also been undertaken by using commercial software FLUENT to validate the thermal plume cases. All of the verified test results closely matched the reconstruction result, which could be evidence of the effectiveness and universality of the method. Additionally, we compared our method with BOS. The advantages of this method can be mainly summarized as high fidelity and low processing time.

Despite the promising results, there are areas for future improvement and exploration, such as the Three-dimensional Effects. During the investigation, the transition part from laminar to turbulence for the hot air jet might bring some uncertainty in the reconstruction. Future work will focus on incorporating three-dimensional effects to enhance the accuracy of the reconstruction. After addressing the three-dimensional effect, more general test cases such as multiple component mixtures and near-sonic jets will be included to validate the universality of the method in the future stage of this research. In conclusion, the physics-informed density field reconstruction method using shadowgraph images presents a significant advancement in flow field diagnostics. It offers a simple, cost-effective, and accurate alternative to traditional measurement techniques. As a surrogate model, it has the potential to revolutionize real-time flow field monitoring and analysis. Although there are several challenges to be overcome, the future of this technique looks promising. It is still believed that this technique could cover and become a reliable tool in various scientific and engineering disciplines.